\documentclass[aps,pra,twocolumn,showpacs,superscriptaddress]{revtex4-1}
\usepackage{graphicx}
\usepackage{dcolumn}
\usepackage{bm}
\usepackage{color}
\usepackage{times}
\usepackage{amssymb}
\usepackage{amsmath}
\usepackage{epsfig}
\usepackage{epstopdf}
\usepackage{dsfont}
\usepackage{subfigure}
\usepackage[colorlinks=true,citecolor=blue, linkcolor=blue,urlcolor=blue]{hyperref}
\usepackage{bookmark}
\usepackage{color}

\definecolor{Green3}{rgb}{0.80,0.87,0.76}

\begin{document}

\title{Quantum mutual information, coherence and unified relations of top quarks in QCD processes }

\author{Duo-Duo Chen}
 \affiliation{School of Physics, Anhui University, Hefei
230601,  People's Republic of China}

\author{Xue-Ke Song}%
 \affiliation{School of Physics, Anhui University, Hefei
230601,  People's Republic of China}

 \author{Liu Ye}
  \affiliation{School of Physics, Anhui University, Hefei
230601,  People's Republic of China}
\author{Dong Wang}
 \email{dwang@ahu.edu.cn}
  \affiliation{School of Physics, Anhui University, Hefei
230601,  People's Republic of China}

\date{\today}

\begin{abstract}
As the most massive particle in the Standard Model, the top quark's exceptionally short lifetime ($\tau \sim 10^{-25} $ s) preserves its spin polarization information through direct decay, making it an ideal system for probing quantum correlations in high-energy physics. In this letter, we presents a comprehensive investigation of quantum correlations in top quark-antiquark ($t\bar{t}$) pairs produced through QCD. We employ multiple quantum information theoretic measures including quantum mutual information, relative entropy of coherence, complete complementarity relations, and the  intrinsic relationship, establishing their dependence on kinematic variables.  Furthermore, we find that for quarks and gluons initial  mixing, as the probability of gluons $\mathcal{W} _{gg}$ increases,  the maximum  of the the left-hand side of the intrinsic relation also increases. We thus believe the current findings are beneficial to insight into the systemic quantumness in QCD.
\end{abstract}

\maketitle

\section{Introduction}
The top quark plays an important role in particle physics, cosmology and other fields. As the most massive fundamental particle known to exist ($m_{t}c^{2} \approx  173$ GeV), it   was  first discovered  by the D0 and CDF collaborations at the Tevatron in the Fermilab \cite{PhysRevD.98.030001,PhysRevLett.74.2632,PhysRevD.50.4517}. In 2008, the Large Hadron Collider (LHC) \cite{BRUNING2012705,Evans_2007}, the highest-energy collider, began operation. The top quark mass, derived from combined Tevatron and LHC Run $1$ and Run $2$ data, is   $m_{t}c^{2} \approx  173.1\pm 0.6$ GeV.   The top quark's large mass results in a correspondingly large decay width, yielding an exceptionally short lifetime $\tau =1/\Gamma _{t}\sim 10^{-25}$ s.  Top quarks are typically produced in top-antitop ($t\bar{t} $) pairs. The process of hadronization and  spin decorrelation cannot affect $t\bar{t}$ spin correlations measurement because of hadronization (with a time scale $\sim 10^{-23}$ ) and spin decorrelation (with a time scale $\sim 10^{-21}$).   This is what makes the top quark so unique and we can reconstruct its spin from the decay products.

The  spin correlation of top quark pairs has been investigated by the D0 and CDF collaborations at the Tevatron with proton-antiproton ($p\bar{p}$) collisions \cite{PhysRevD.83.031104,PhysRevLett.107.032001}, and by the ATLAS and CMS collaborations at the LHC with proton-proton ($pp$) collisions \cite{PhysRevLett.114.142001,PhysRevD.100.072002,PhysRevD.93.012002,2016321}. In a particle accelerator, two beams of protons (or antiprotons and  protons) are accelerated in opposite directions and collide with each other. These collisions produce large number of particles, including   the  top quark and anti-top quark pairs. A proton consists of two up quarks (spin-1/2 fermions) and one down quark. Quarks-elementary particles that interact via the strong force-are the fundamental constituents of matter. They are tightly bound together by gluons (massless spin-1 bosons) to form stable protons. An antiprotons is the antimatter counterpart of the proton. It is a composite particle consisting of two anti-up quarks and one anti-down quark. Antiprotons can remain stable in high-vacuum isolation, but upon contact with ordinary matter (e.g., protons), they almost instantaneously annihilate, converting their mass into other particles  and energy.

Establishing a connection between high-energy physics and quantum information science is extremely important.    Afik \emph{et al.} proposed the detection of entanglement between the spins of top-antitop-quark pairs at the LHC, representing the  entanglement observation at the highest energy scale so far in 2021 \cite{Afik_2021}, computed the quantum state of a $t\bar{t} $ pairs produced from the most elementary quantum chromodynamics (QCD) processes, finding the presence of entanglement and CHSH violation in different regions of phase space in 2022 \cite{Afik2022quantuminformation} and also provide the full picture of quantum correlations in top quarks by studying also quantum discord and steering in 2023 \cite{PhysRevLett.130.221801}. In 2024, Ye  \emph{et al.} employed the entropic uncertainty relations and the quantum Fisher information to explore the formation of quark $t\bar{t}$ pairs at the LHC through the combination of $q\bar{q} $ pairs and $gg$ pairs initiated processes \cite{PhysRevD.110.055025}. In 2025, Cheng \emph{et al.} show analytically that the basis that diagonalizes the spin-spin correlations is optimal for maximizing spin correlations, entanglement, and Bell inequality violation  \cite{PhysRevD.111.033004}. Furthermore,  Han \emph{et al.} employ two complementary approaches for the study of the top anti-top system, namely the decay method and the kinematic method and they highlight subtleties associated with measuring discord for reconstructed quantum states at colliders in 2025 \cite{JHEP05(2025)081}.

In this work, we present a comprehensive investigation of quantum correlations in top quark-antiquark ($t\bar{t}$) pairs produced via QCD processes. Rather than being restricted to specific collider settings, we perform a general theoretical scan over the full range of initial-state mixtures ($w_{gg} \in [0,1]$) to reveal the universal scaling of quantum observables with production mechanisms. We apply several key information theoretic tools---namely quantum mutual information (QMI),  relative entropy of coherence (REC), complete complementarity relation (CCR), and an intrinsic relation to quantify and characterize the entanglement structure within the $t\bar{t}$ system at the QCD. These observables go beyond conventional spin correlation measures by capturing both quantum and classical correlations in a unified framework. Specifically, QMI quantifies the total correlations between the top and antitop quarks. As a standard measure in quantum resource theory, REC quantifies the quantum coherence of a state.  The CCR provides a complementary relation that links coherence, predictability, and correlations, revealing how quantum information is distributed among subsystems. The intrinsic relation further connects uncertainty, coherence, and predictability, offering a more complete picture of the quantumness of the system.

Unlike traditional observables such as spin correlations or entanglement witnesses, these quantities are continuous values, typically ranging from 0 to $2\log_{2}{d}$ for QMI and from 0 to $\log_{2}{d}$ for REC in qubit system ($d$ is the dimension of a single particle). A larger value of QMI indicates the stronger total correlations including classical and quantum parts, while a larger REC signifies the greater quantum coherence (i.e., quantumness). Importantly, these observables allow us to distinguish between different types of quantum behavior much as coherence versus entanglement and to explore how they vary across kinematic phase space. By measuring them in $t\bar{t}$ pair production, we gain new insights into the structure of quantum correlations in high-energy collisions. Thereby, the QMI and REC can intrinsically reveal the global correlation and quantumness in the course of particle collisions.

The Letter is structured as follows. In Sect. \ref{sect2}, we  review  in detail the production of $t\bar{t} $ pairs through  QCD processes.  In Sect. \ref{sect3}, we discuss the QMI in $t\bar{t} $ production.   In Sect. \ref{sect4}, we  study the  relative entropy of coherence  in $t\bar{t} $ production. In Sect. \ref{sect5}, we  investigate the CCR   among   QMI, the conditional entropy, REC, and predictability  in $t\bar{t} $ pairs. In Sect. \ref{sect6}, we explore the  intrinsic relation among entropic uncertainty, REC and predictability in $t\bar{t} $ pairs. Finally, we end up our Letter with a concise conclusion in Sect. \ref{sect7}.

\section{Preliminaries of $t\bar{t}$ pairs production}\label{sect2}
A natural Standard Model candidate for an entangled two-qubit system is a particle-antiparticle (denoted generically as $P\bar{P} $) pairs produced from some initial state $I_{s}$
\begin{align}
I_{s}\to P+\bar{P},
 \label{Eq.2}
    \end{align}
in this paper, we only discuss the case where $P$ is a  fermion of mass $m$ in the Standard Model. The components of the initial state $I_{s}$ annihilate themselves through interactions, resulting in    the the production of $P\bar{P} $ pairs. In this paper, we study particle-antiparticle pairs ($P\bar{P} $), focusing on top-antitop quark pairs ($t\bar{t} $) produced via QCD processes in high-energy colliders. In the theoretical calculation, we employ  leading-order (LO) QCD perturbation theory.

The interactions between a light quark and antiquark ($q\bar{q} $) or a pair of gluons ($gg$) through QCD result in the production of a $t\bar{t} $ pair as follows
\begin{align}
\begin{split}
  &q +  \bar{q}\to t+  \bar{t} \\
 &g +  g\to t+  \bar{t},
 \label{Eq.3}
    \end{split}
    \end{align}
the production of a $t\bar{t} $ pair is described by the invariant mass $M_{t\bar{t}}$  and the direction $\hat{k } $ in the center-of-mass frame (c.m.). In this frame, the top and antitop relativistic momenta are $k_{t}^{ \mu    } = (k_{t}^{0    },\mathrm{\mathbf{k} } )$, $k_{\bar{t} }^{ \mu    } = (k_{\bar{t} }^{0    },\mathrm{\mathbf{-k} })$, satisfying the invariant dispersion relation $k_{t }^{ 2   }\equiv k_{t }^{ \mu }k_{\mu t}  = m_{t}^{2}$, and similar for the antitop  $k_{\bar{t} }^{2    }=k_{t }^{2}=m_{t}^{2}$.

The invariant mass $M_{t\bar{t} }$ in the c.m. energy is defined as
\begin{align}
M_{t\bar{t} }^{2} \equiv s_{t\bar{t} }\equiv(k_{t}+k_{ \bar{t}}  ),
 \label{Eq.4}
    \end{align}
 in the c.m. frame, $M_{t\bar{t} }^{2}=4(k_{t }^{0    }  )^{2} =4(m_{t}^{2}+\mathrm{\mathbf{k} }^{2}  )$, the momentum of top quark is related to its velocity $\beta$, satisfying $\left | \mathrm{\mathbf{k} }  \right | =m_{t}\beta /\sqrt{1-\beta ^{2} } $, we obtain
 \begin{align}
\beta = \sqrt{1-4m_{t}  ^{2}/M_{t\bar{t} }^{2}  }.
 \label{Eq.5}
    \end{align}
The production threshold for $t\bar{t}$ pairs, corresponding to $\beta = 0$, occurs at  the lowest energy, $M_{t\bar{t} }=2m_{t}\approx 346$ GeV.

The spin correlations of a $t\bar{t}$ pair are fully described, at a given partonic center-of-mass energy and scattering angle, by the so-called production spin density matrix $\tilde{\rho}$. In the basis of tensor products of top-quark and anti-top-quark spin states, the density matrix can be parameterized as
\begin{align}
\tilde{\rho}=\tilde{A} \mathcal{I}_4 +\sum_{i}^{} (\tilde{B}_{i}^{+} \sigma _{i} \otimes   \mathcal{I}_2 +\tilde{B}_{i}^{-}\mathcal{I}_2\otimes \sigma _{i}    )+\sum_{i,j}^{}\tilde{C}_{ij}\sigma _{i} \otimes \sigma _{j}.
 \label{Eq.6}
    \end{align}
The production spin density matrix $\tilde{\rho}$ is characterized by 16 parameters, $\tilde{A}$, $\tilde{B}_{i}^{\pm }$, $\tilde{C}_{ij}$, with $\tilde{A}$ determining the differential cross section for $t\bar{t} $ production at fixed energy and top direction.

The proper spin density matrix $\hat{\rho}$ of the  $t\bar{t} $ pairs are  obtained from  normalizing $\tilde{\rho}$
\begin{align}
\hat{\rho}=\frac{\tilde{\rho}}{\operatorname{Tr}(\tilde{\rho})} =\frac{\tilde{\rho}}{4\tilde{A} },
 \label{Eq.7}
    \end{align}
where $\tilde{A}$   represents the differential cross-section for the production of $t\bar{t} $ at a fixed energy and in a specified top-quark direction.

As a result, the spin polarizations $B_{i}^{\pm }$ and spin correlations $C_{ij} $ of the $t\bar{t}$ pairs are normalized by the total cross section coefficient $\tilde{A}$
\begin{align}
B_{i}^{\pm } =\frac{\tilde{B}_{i}^{\pm } }{\tilde{A} },C_{ij}=\frac{\tilde{C}_{ij}^{\pm }} {\tilde{A} }.
 \label{Eq.8}
    \end{align}

 The spin density matrix $\hat{\rho}$ in the helicity  basis is then given by
 \begin{align}
\hat{\rho} =\frac{1}{4}(  \mathcal{I}_4 +\sum_{i}^{} (B_{i}^{+} \sigma _{i} \otimes \mathcal{I}_2+B_{i}^{-}\mathcal{I}_2\otimes \sigma _{i}    )+\sum_{i,j}^{}C_{ij}\sigma _{i} \otimes \sigma _{j} ).
 \label{Eq.9}
    \end{align}
In this framework, the correlation matrix $C_{ij}$  and spin polarizations $B_{i}^{\pm }$ are constrained by the LO symmetries. The normalized form of the spin density matrix $\hat{\rho}$, expressed in the helicity basis, is fully characterized by five independent parameters ($\tilde{C}_{rr}, \tilde{C}_{nn}, \tilde{C}_{kk}, \tilde{C}_{rk}, \tilde{C}_{kr}$), where $\hat{\rho}$ depends only on  $\beta = \sqrt{1-4m_{t}  ^{2}/M_{t\bar{t} }^{2}  }$ and $\Theta$.  At LO, in the Standard Model, the $t\bar{t}$ pairs is unpolarized ($\tilde{B}_{i}^{\pm }=0$), and the correlation matrix satisfies $\tilde{C}_{ij}=\tilde{C}_{ji}$. The explicit form of $\hat{\rho}$ is given by
\begin{align}
\begin{split}
\hat{\rho} =\frac{1}{4\tilde{A} }\begin {pmatrix}
\tilde{A}+\tilde{C}_{kk}  & \tilde{C}_{kr} & \tilde{C}_{rk} & \tilde{C}_{rr}-\tilde{C}_{nn}\\
 \tilde{C}_{kr} & \tilde{A}-\tilde{C}_{kk} & \tilde{C}_{rr}+\tilde{C}_{nn} & -\tilde{C}_{rk}\\
 \tilde{C}_{rk} & \tilde{C}_{rr}+\tilde{C}_{nn} & \tilde{A}-\tilde{C}_{kk} & -\tilde{C}_{kr}\\
 \tilde{C}_{rr}-\tilde{C}_{nn} &  -\tilde{C}_{rk}& -\tilde{C}_{kr} &\tilde{A}+\tilde{C}_{kk}
\end{pmatrix}.
\label{Eq.10}
    \end{split}
    \end{align}

For $q\bar{q} $ process, the coefficients of the  matrix are
\begin{align}
\begin{split}
\tilde{A}^{q\bar{q} }&=F_{q}(2-\beta ^{2}\sin^{2}\Theta   ), \\
\tilde{C}_{rr}^{q\bar{q} }&=F_{q}(2-\beta ^{2})\sin^{2}\Theta, \\
\tilde{C}_{nn}^{q\bar{q} }&=-F_{q}\beta ^{2}\sin^{2}\Theta, \\
\tilde{C}_{kk}^{q\bar{q} }&=F_{q}[2-(2-\beta ^{2})\sin^{2}\Theta ], \\
\tilde{C}_{rk}^{q\bar{q} }&=\tilde{C}_{kr}^{q\bar{q} }=F_{q}\sqrt{1-\beta ^{2}} \sin2\Theta, \\
F_{q}&=\frac{1}{18}.
 \label{Eq.11}
    \end{split}
    \end{align}

For $gg$ process, the coefficients of the matrix are
\begin{align}
\begin{split}
\tilde{A}^{gg }&=F_{g}[1+2\beta ^{2}\sin^{2}\Theta-\beta ^{4}(1+\sin^{4}\Theta)   ], \\
\tilde{C}_{rr}^{gg }&=-F_{g}[1-\beta ^{2}(2-\beta ^{2})(1+\sin^{4}\Theta)], \\
\tilde{C}_{nn}^{gg}&=-F_{g}[1-2\beta ^{2}+\beta ^{4}(1+\sin^{4}\Theta)], \\
\tilde{C}_{kk}^{gg}&=-F_{g}[1-\beta ^{2}\frac{\sin^{2} 2\Theta }{2} -\beta ^{4}(1+\sin^{4}\Theta) ], \\
\tilde{C}_{rk}^{gg}&=\tilde{C}_{kr}^{gg}=F_{g}\sqrt{1-\beta^{2} } \beta ^{2}\sin2\Theta\sin^{2}\Theta, \\
F_{g}&=\frac{7+9\beta ^{2}\cos^{2}\Theta   }{192{(1-\beta ^{2}\cos^{2}\Theta)^2} }.
 \label{Eq.12}
    \end{split}
    \end{align}

The spin density matrix for the mixed state of $q\bar{q} $ and $gg$ initial states can be expressed as
\begin{align}
\hat{\rho}(M_{t\bar{t},\hat{k}  } )=\sum_{I=q\bar{q},gg}\mathcal{W }_{I} (M_{t\bar{t},\hat{k}  } )\hat{\rho}^{I} (M_{t\bar{t},\hat{k}  } ).
 \label{Eq.13}
    \end{align}
In this work, we treat $\mathcal{W }_{gg}$ as a free parameter ranging from 0 to 1, allowing us to systematically explore how the quantum nature of $t\bar{t}$ pairs depend on the initial-state composition. This parameter covers the full range of physical scenarios: from low-energy colliders where $q\bar{q}$ annihilation dominates (e.g., Tevatron, $\mathcal{W }_{gg}\approx 0.1$), to high-energy colliders where $gg$ fusion dominates (e.g., LHC at 13 TeV, $\mathcal{W }_{gg}\approx 0.85$--$0.95$).


\section{The quantum mutual information in $t\bar{t} $}\label{sect3}
In   quantum information theory, QMI provides a quantitative measure of correlations between two quantum systems \cite{PhysRevE.69.066138,PhysRevA.72.032317,shin2024estimating,PhysRevA.74.042305}. As a generalization of classical mutual information, QMI serves as a fundamental tool for quantifying both quantum and classical correlations, including entanglement  \cite{RevModPhys.81.865,GUHNE20091} that exist between the  systems. In a two-qubit mixed state, QMI ranges from $0$ (uncorrelated) to  $2\log_{2}{d}$ (maximally correlated), with the maximum achieved by states of either maximal classical correlation or maximal entanglement. The  QMI quantifies the total correlation between subsystems: a larger value indicates stronger total correlation.   Given an arbitrary bipartite quantum state   $\hat{\rho} _{AB}$, the QMI is formally defined by \cite{PhysRevA.72.032317}
\begin{align}
\begin{split}
I(A:B)&=S(\hat{\rho} _{A} )+S(\hat{\rho} _{B} )-S(\hat{\rho} _{AB} )\\
&=S(\hat{\rho} _{A} )-S(A|B).
 \label{Eq.14}
    \end{split}
    \end{align}
 The von Neumann entropy, denoted as $S(\hat{\rho} _{ } )$,   quantifies the information content of a quantum state $\hat{\rho}$ in a d-dimensional Hilbert space. It is formally defined as  $S(\hat{\rho} )=-\operatorname{Tr}(\hat{\rho} \log_{}{\hat{\rho}} )$. For a bipartite quantum system described by the state $\hat{\rho} _{AB}$, the conditional entropy $S(A|B)$ prior to measurement is given by the difference $S(\hat{\rho} _{AB})-S(\hat{\rho} _{B})$,  where $\hat{\rho} _{B}=\operatorname{Tr} _{A}(\hat{\rho} _{AB})$ represents the reduced state of subsystem $B$. The detailed mathematical expression for QMI is provided in \hyperref[app:rec]{Appendix~\ref*{app:rec}}.
\begin{figure}
\begin{minipage}{0.5\textwidth}
\centering
\subfigure{\includegraphics[height=3.7cm]{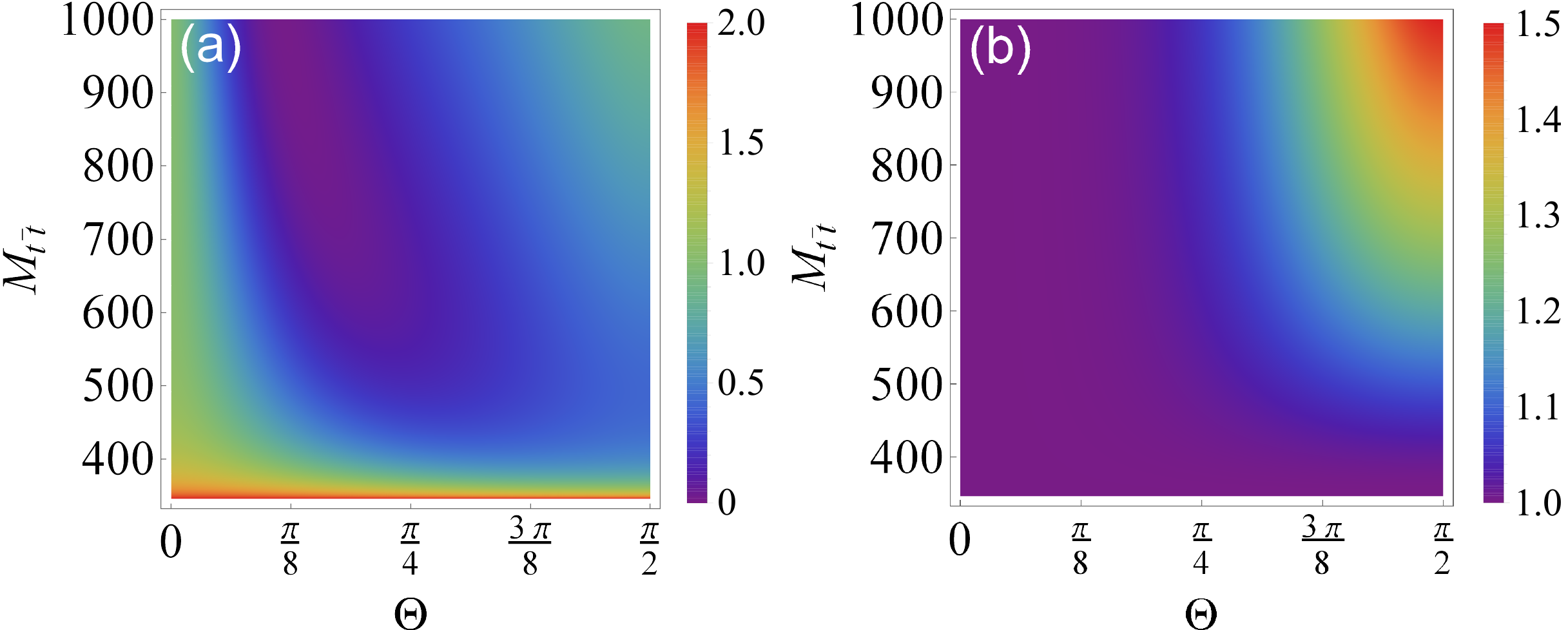}}
\end{minipage}\hfill
    \caption{QMI as a function of  invariant mass $M_{t\bar{t} }$ and  production angle $\Theta$ for   $t\bar{t}$ pairs production processes. (a): $gg \to t\bar{t}$; (b): $q\bar{q} \to t\bar{t}$.}
    \label{f1}
  \end{figure}

Fig. \ref{f1} shows the QMI analysis for $t\bar{t}$ pairs production at the QCD.  Fig. \ref{f1}(a) displays the results for  the gluon fusion channel ($gg \to t\bar{t}$). where $I(A:B)$ exhibits dependence on both production angle $\Theta$ and invariant mass
$M_{t\bar{t} }$. The QMI demonstrates a gradual decrease with increasing $\Theta$,  with maximal entanglement observed near the threshold region at $M_{t\bar{t} }\approx 346$ GeV. Fig. \ref{f1}(b) displays the quark annihilation channel ($q\bar{q} \to t\bar{t}$),  where QMI reaches a minimum near threshold and increases with $M_{t\bar{t} }$ as $\Theta$ increases. The peak QMI occurs at $M_{t\bar{t} }=1000$ GeV and $\Theta=\pi/2$, with the system maintaining both classical and quantum correlations throughout the kinematic range.

\begin{figure}
\begin{minipage}{0.5\textwidth}
\centering
\subfigure{\includegraphics[height=7.5cm]{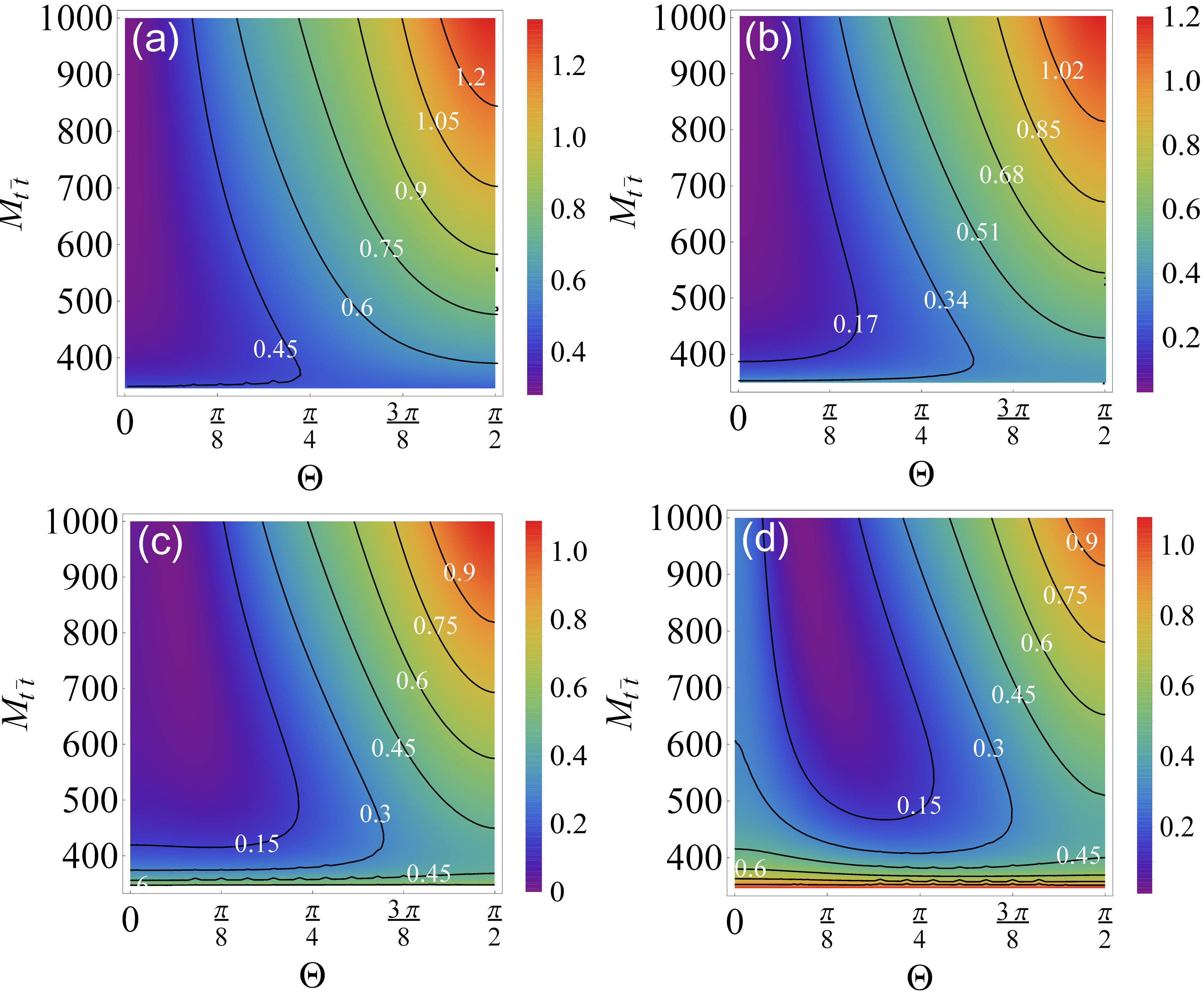}}
\end{minipage}\hfill
    \caption{QMI in $t\bar{t}$ production with mixed gluons $(gg)$ and quarks $(q\bar{q})$ initial state, where the gluons   probability $\mathcal{W} _{gg}$ varies:
    (a): $\mathcal{W} _{gg}=0.2$; (b):  $\mathcal{W} _{gg}=0.4$; (c):  $\mathcal{W} _{gg}=0.6$; (d):  $\mathcal{W} _{gg}=0.8$. }
    \label{f2}
  \end{figure}
Fig. \ref{f2} presents a systematic study of quantum correlations in $t\bar{t}$ production using a simplified model with fixed partonic probabilities. The analysis examines how varying the mixture between $gg$ and $q\bar{q}$ channels affects the quantum state $\hat{\rho}(M_{t\bar{t},\hat{k}  } )$. QMI is shown as a function of production angle $\Theta$ and the invariant mass $M_{t\bar{t} }$, with subfigures (a)-(d) corresponding to increasing gluon fusion probabilities $\mathcal{W} _{gg}=  0.2, 0.4, 0.6,$ and $0.8$ respectively. At low $\mathcal{W} _{gg}$, maximal $I(A:B)$ occurs in the large mass, large-angle region (upper right corner). Notably, as $\mathcal{W} _{gg}$ increases to $0.8$, the QMI behavior converges toward the pure $gg\to t\bar{t}$ case shown in Fig. \ref{f1}(a).

\section{The relative entropy of coherence in $t\bar{t} $}\label{sect4}

Quantum coherence, a fundamental manifestation of the superposition principle in quantum mechanics, has evolved into a key resource for quantum technologies \cite{PhysRev.131.2766,PhysRevLett.113.140401,PhysRevA.68.053802,PhysRevA.92.042101,PhysRevA.91.042133,PhysRevA.106.042415,PhysRevA.110.032420,PhysRevA.110.012418,streltsov2017colloquium}. Originating from the ability of quantum states to exist in superpositions, coherence enables diverse quantum-enhanced applications. The quantification of coherence has been rigorously formalized \cite{PhysRevLett.115.020403,PhysRevLett.116.150502}, with two principal measures emerging: the $l_{1}$-norm coherence and the REC. For a two-particle mixed state $\hat{\rho}_{AB}$, the REC $C_{\text{re}}(\hat{\rho}_{AB})$ satisfies $0<C_{\text{re}}(\hat{\rho}_{AB})<  \log_{2}{d}$, where $d$ is the dimension of a single particle. A larger value of $C_{\text{re}}(\hat{\rho}_{AB})$ indicates stronger quantum coherence in the state. The  REC $C_{\text{re}}(\hat{\rho}_{AB})$ is defined mathematically as \cite{PhysRevLett.113.140401}
\begin{align}
C_{\text{re}}(\hat{\rho}_{AB})=S(\hat{\rho} _{diag} )-S(\hat{\rho}_{AB} ),
\label{Eq.15}
\end{align}
where $S(\hat{\rho}_{AB})=-\operatorname{Tr}(\hat{\rho}_{AB} \log_{}{\hat{\rho}_{AB} })$ denotes the von Neumann entropy, and $\hat{\rho} _{diag} $ is the non-coherence state and obtained by taking diagonal elements of the matrix $\hat{\rho}  _{AB}$.

\begin{figure}
\begin{minipage}{0.52\textwidth}
\centering
\subfigure{\includegraphics[height=3.7cm]{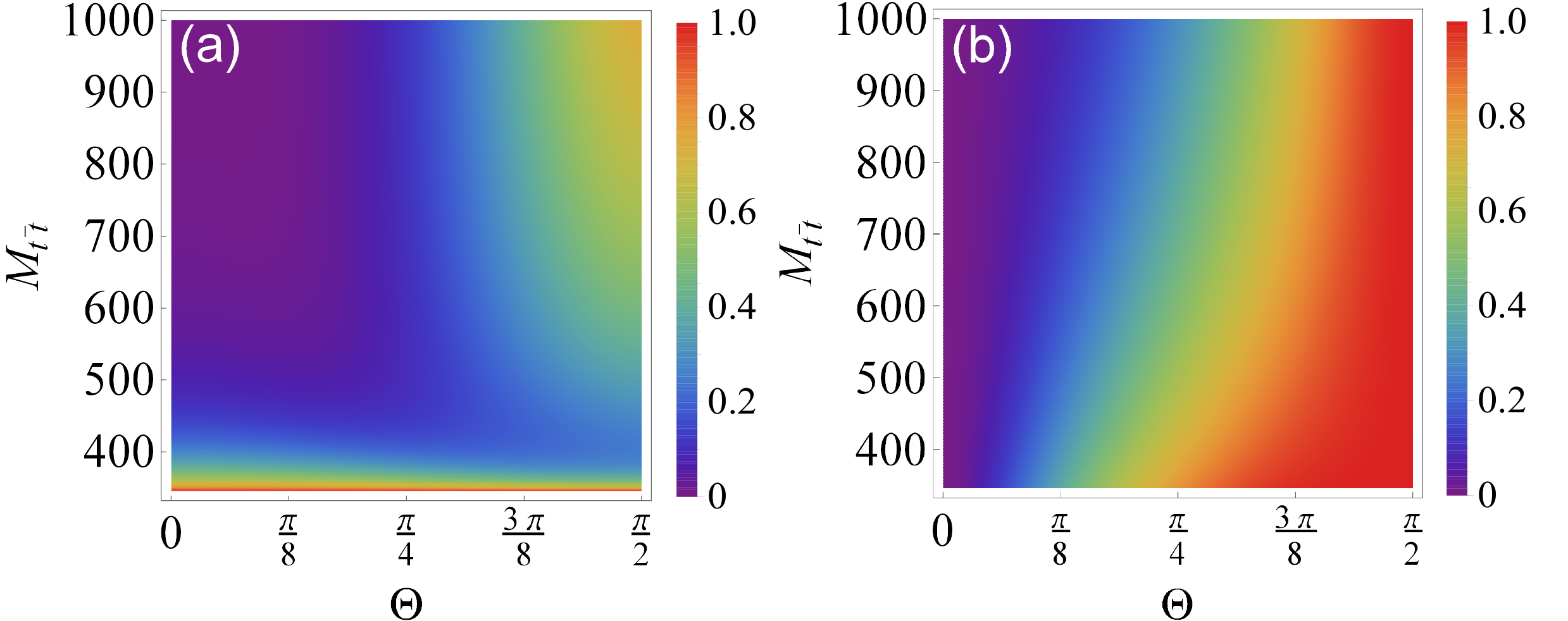}}
\end{minipage}\hfill
    \caption{REC as a function of the invariant mass $M_{t\bar{t} }$ and the production angle $\Theta$ in a $t\bar{t}$ pairs. (a): $gg \to t\bar{t}$. (b): $q\bar{q} \to t\bar{t}$.}
    \label{f3}
  \end{figure}

The REC for the system is derived from Eqs. (\ref{Eq.11}), (\ref{Eq.12}) and (\ref{Eq.15}). The detailed mathematical expression for REC is provided in \hyperref[app:rec]{Appendix~\ref*{app:rec}}. In Fig. \ref{f3}(a), we present the REC for $t\bar{t}$ production via gluon fusion ($gg \to t\bar{t}$), showing its dependence on both the production angle  $\Theta$ and the invariant mass $M_{t\bar{t} }$.  Notably, when $M_{t\bar{t} }$ is near $346$ GeV, $C_{\text{re}}(\hat{\rho}_{AB})$ decreases monotonically with increasing $\Theta$. In contrast, Fig. \ref{f3}(b) displays the REC for $t\bar{t}$ production through quark-antiquark annihilation ($q\bar{q} \to t\bar{t}$). Here, $C_{\text{re}}(\hat{\rho}_{AB})$  increases monotonically with $\Theta$  at fixed $M_{t\bar{t} }$.

\begin{figure}
\begin{minipage}{0.5\textwidth}
\centering
\subfigure{\includegraphics[height=7.52cm]{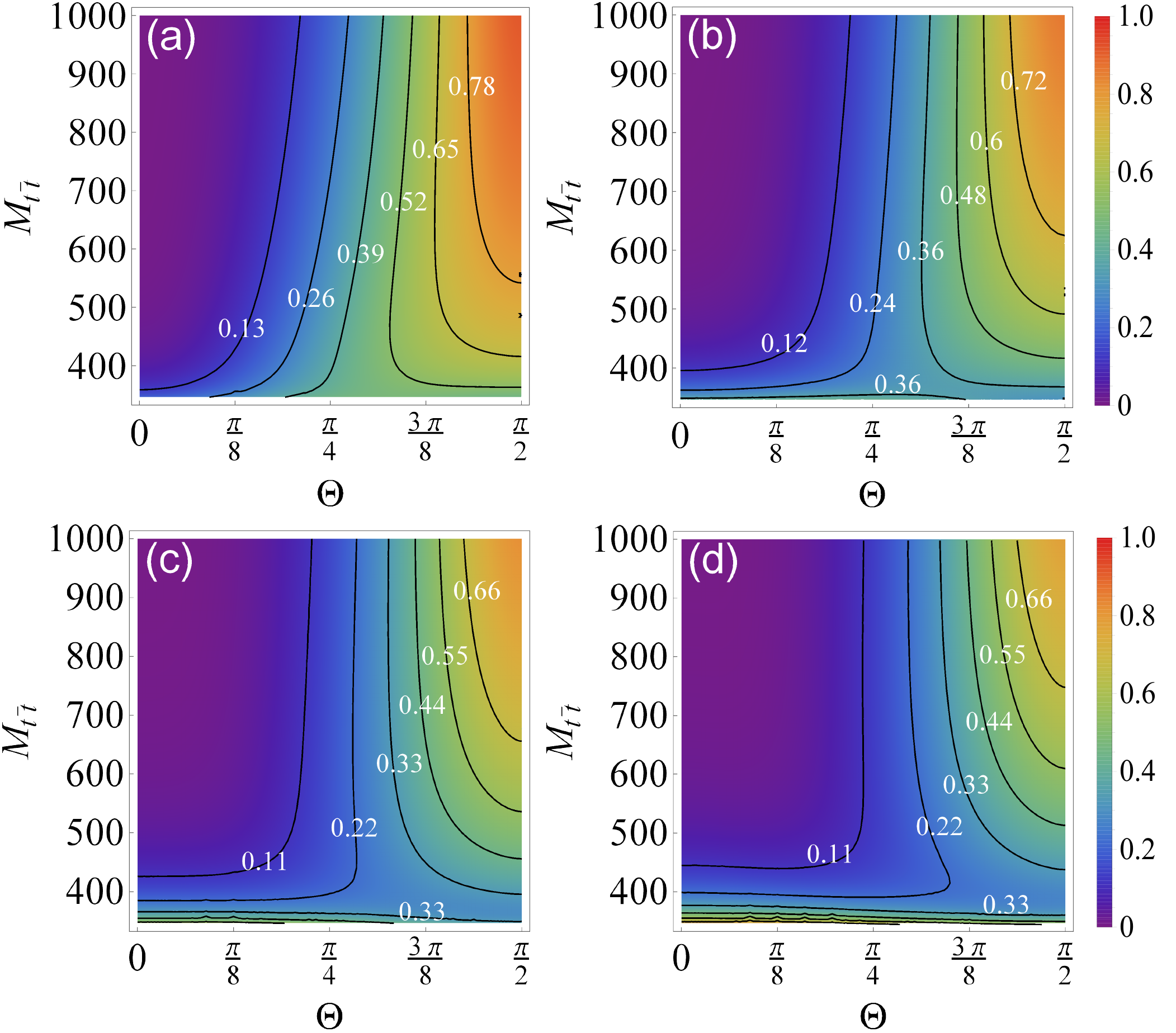}}
\end{minipage}\hfill
    \caption{REC $C_{\text{re}}(\hat{\rho}_{AB})$ in $t\bar{t}$ production for mixed gluons $(gg)$ and quarks $(q\bar{q})$ initial state  with varying gluon probability $\mathcal{W} _{gg}$: (a): $\mathcal{W} _{gg}=0.2$; (b):  $\mathcal{W} _{gg}=0.4$; (c):  $\mathcal{W} _{gg}=0.6$; (d):  $\mathcal{W} _{gg}=0.8$.}
    \label{f4}
  \end{figure}

Fig. \ref{f4}  displays the REC  $C_{\text{re}}(\hat{\hat{\rho}}_{AB})$ for   $t\bar{t}$  production  through both quark annihilation   ($q\bar{q} \to t\bar{t}$) and gluons fusion ($gg \to t\bar{t}$)  channels. Subfigures (a)-(d) correspond to gluon fusion probabilities $\mathcal{W} _{gg}=0.2, 0.4, 0.6,$ and $0.8$, respectively, showing the REC dependence on production angle $\Theta$ and invariant mass $M_{t\bar{t} }$. With increasing $\mathcal{W} _{gg}$, we observe that suppression of low quality/low angle on REC  is weakening. REC  shrinks in the right corner range  and   expands in the left corner range in Fig. \ref{f4}(a)-(d).

\section{The complete complementarity relations  in $t\bar{t} $}\label{sect5}
The CCR \cite{bittencourt2024quantifying,Basso_2020} provides a rigorous framework for quantifying quantum correlation in the composite system. We apply this approach to characterize quantum correlations in  $t\bar{t} $ pairs  production at  the QCD. The CCR are constructed from von Neumann entropy-based measures of predictability and coherence. For bipartite mixed states, the  CCR  takes the form \cite{bittencourt2024quantifying}
\begin{align}
\begin{split}
  &I_{A:B}(\hat{\rho} _{AB} )+S_{A|B}(\hat{\rho} _{AB} ) +P_{\text{vn}}(\hat{\rho} _{A} )+C_{\text{re}}(\hat{\rho} _{A} ) \\
 &=\log_{2}{d_{A} }.
 \label{Eq.1}
    \end{split}
    \end{align}
QMI $I_{A:B}(\hat{\rho} _{AB} )$  quantifies the correlations between subsystems $A$ and $B$. The conditional von Neumann entropy $S_{A|B} (\hat{\rho} _{AB} )= S_{\text{vn}} (\hat{\rho} _{AB} )-S_{\text{vn}} (\hat{\rho} _{B} )$ quantifies the uncertainty in subsystem
$A$ conditioned on subsystem $B$. For a d-dimensional Hilbert space,  $S_{A|B} (\hat{\rho} _{AB} )$ is bounded as $-\log_{2}{d} \leqslant S_{A|B} (\hat{\rho} _{AB} )\leqslant \log_{2}{d}$ ($d$ is the dimension of subsystem $A$ or $B$). A positive conditional entropy ($S_{A|B} (\hat{\rho} _{AB} )>0$) suggests either a classical state or a quantum state with weak correlations, whereas a negative value ($S_{A|B} (\hat{\rho} _{AB} )<0$) guarantees the presence of quantum entanglement. Here, $S_{\text{vn} } (\hat{\rho} _{})$ denotes the von Neumann entropy of the state $\hat{\rho} _{}$, and $\hat{\rho} _{A, \text{diag}}={\textstyle \sum_{i=1 }^{d_{A} }\hat{\rho} _{ii}^{A} \left| i \right\rangle \left\langle i \right|  }$ represents the diagonalized reduced state of subsystem $A$. Additionally, the predictability measure $P_{\text{vn} } (\hat{\rho} _{A} )$, defined as $P_{\text{vn} } (\hat{\rho} _{A} )\equiv \log_{2}{d_{A} } -S_{\text{vn} } (\hat{\rho} _{A},\text{diag}  )$, denotes the predictability of the single subsystem  $A$.

When considering the state of a single subsystem $A$,   information shared through its correlations
with subsystem $B$ is inevitably lost.  The  reduced density matrix of subsystem $A$ can be expressed as $\hat{\rho }_{A} =\frac{1}{2}\mathcal{I}_2$. $\hat{\rho }_{A}$ in  a two-dimensional Hilbert space.

From Eq. (\ref{Eq.1}) and $\hat{\rho }_{A} =\frac{1}{2}\mathcal{I}_2$, we  find that  both the predictability measure $P_{\text{vn}}(\hat{\rho} _{A} )$ and the coherence measure $C_{\text{re}}(\hat{\rho} _{A} )$  vanish (become zero). Consequently,  the CCR for $t\bar{t} $ production reduce to functions solely dependent on the  QMI  $I_{A:B}(\hat{\rho} _{AB} )$ and the conditional entropy $S_{A|B}(\hat{\rho} _{AB} )$.
 \begin{figure}
\begin{minipage}{0.5\textwidth}
\centering
\subfigure{\includegraphics[height=7.42cm]{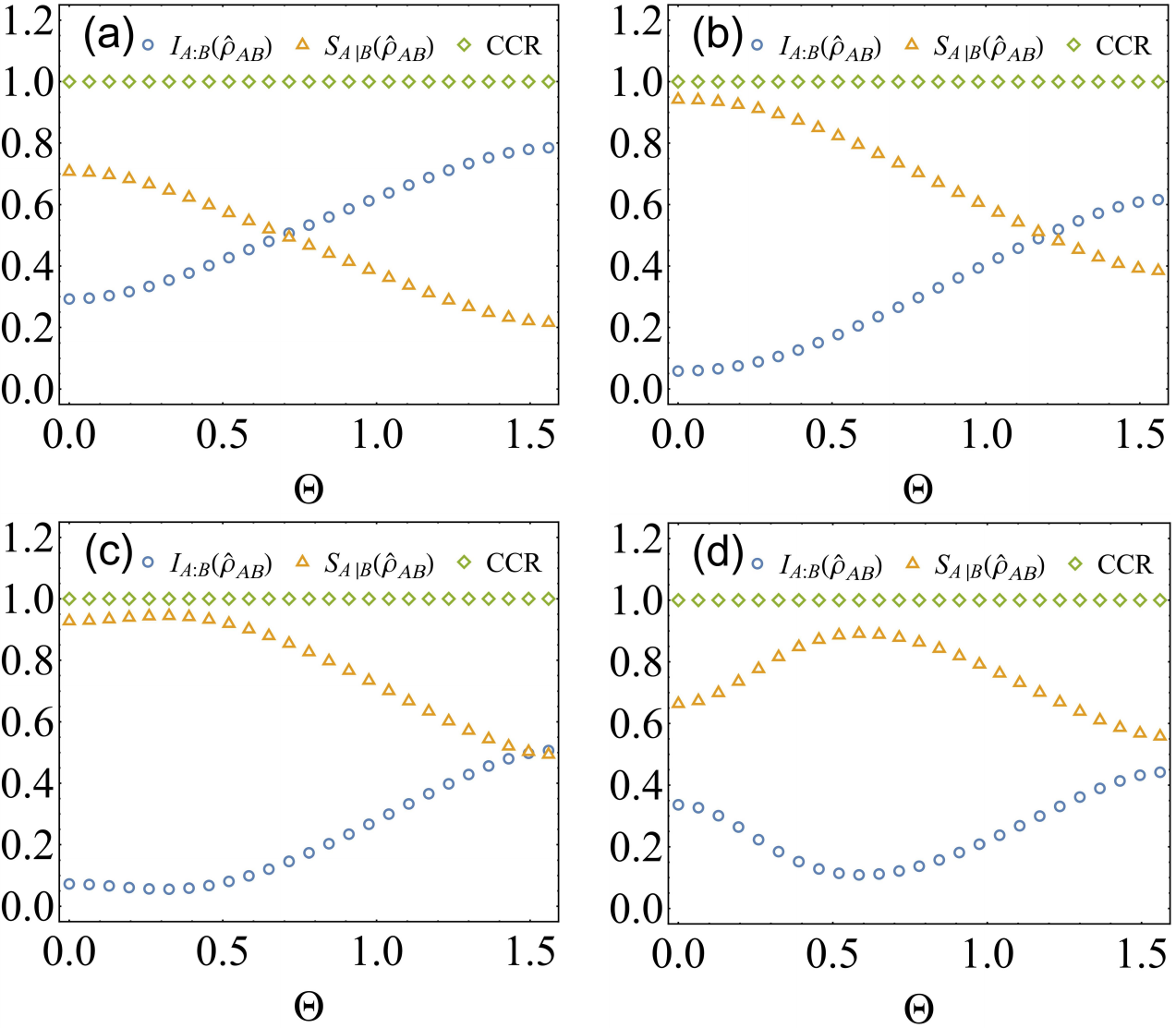}}
\end{minipage}\hfill
    \caption{CCR in $t\bar{t}$ production from mixed initial states of gluons $(gg)$ and quarks $(q\bar{q})$. Blue circles represent the QMI $I_{A:B}(\hat{\rho} _{AB} )$, orange triangles show conditional entropy $S_{A|B}(\hat{\rho} _{AB} )$, and  green diamonds indicate the CCR. The gluons probability $\mathcal{W} _{gg}$ varies: (a): $\mathcal{W} _{gg}=0.2$; (b):  $\mathcal{W} _{gg}=0.4$; (c):  $\mathcal{W} _{gg}=0.6$; (d):  $\mathcal{W} _{gg}=0.8$. The invariant mass is fixed to $M_{t\bar{t} }=500$ GeV.}
    \label{f5}
  \end{figure}

Fig. \ref{f5} shows the dependence of QMI $I_{A:B}(\hat{\rho} _{AB} )$ and conditional entropy $S_{A|B } (\hat{\rho} _{AB} )$ on both the production angle $\Theta$ and the invariant mass $M_{t\bar{t} }$. As evident from the figure, these quantities exhibit complementary behavior: $I_{A:B}(\hat{\rho} _{AB} )$ increases when $S_{A|B } (\hat{\rho} _{AB} )$  decreases, and vice versa. Notably, their sum remains constant at unity ($I_{A:B}(\hat{\rho} _{AB} )+S_{A|B} (\hat{\rho} _{AB} )=1$), demonstrating conservation that is independent of the initial-state mixing between gluons $(gg)$ and quarks $(q\bar{q})$ production channels.

 From Fig. \ref{f5}, we can know that the sum of  $I_{A:B}(\hat{\rho} _{AB} )$ and $S_{A|B } (\hat{\rho} _{AB} )$ remains  conserved at unity ($I_{A:B}(\hat{\rho} _{AB} )+S_{A|B} (\hat{\rho} _{AB} )=1$), independent of  the initial-state mixing between gluons and quarks channels.

\section{The intrinsic relation  in $t\bar{t} $}\label{sect6}
In quantum resource theory, CCR   provide   a powerful framework for characterizing and quantifying quantum correlations and their interplay. As shown in \cite{bittencourt2024quantifying}, the generalized   CCR for bipartite mixed states provides a comprehensive framework   to analyzing quantum spin correlations in top quark pairs production.  Building upon this foundation, we now investigate these quantum correlations in $t\bar{t}$ production  within the QCD framework through the formalism of von Neumann entropy, examining their fundamental relationships in  entropy space \cite{renes2009conjectured,berta2010uncertainty,PhysRevA.110.062220,PhysRevA.106.062219,WANG2024138876}.

For a two-qubit  system,  the measure of predictability $P_{\text{vn}}( \hat{\rho}_{AB}  )$,   REC $C_{\text{re}}(\hat{\rho}_{AB})$,  conditional entropy $S(A|B)$ of the state $\hat{\rho}_{AB}$ and entropic uncertainty relations can be defined as
 \begin{align}
\begin{split}
&P_{\text{vn}}( \hat{\rho}_{AB}  ) =\log_{2}({d_{A}d_{B} } )-S( \hat{\rho}_{d} ),\\
&C_{\text{re}}(\hat{\rho}_{AB})=S(\hat{\rho} _{diag} )-S(\hat{\rho}_{AB} ),\\
&S(A|B)=S(\hat{\rho }_{AB}  )-S(\hat{\rho }_{B}  ),\\
&S(\hat{Q} \left | B \right  )+S(\hat{R} \left | B \right  )\geqslant\log_{2}\frac{1}{c}+ S(A|B).
\label{Eq.17}
    \end{split}
    \end{align}

From Eq. (\ref{Eq.17}), we derive for a two-qubit mixed state system the intrinsic relation  among  entropic uncertainty, REC, and predictability, expressed in terms of von Neumann entropy as
\begin{align}
\begin{split}
&S(\hat{Q} \left | B \right  )+S(\hat{R} \left | B \right  )+S( \hat{\rho}_{B}  )+P_{\text{vn}( \hat{\rho}_{AB}  )}+C_{\text{re}}(\hat{\rho}_{AB})\\ &\geqslant\log_{2}\frac{1}{c}+ \log_{2}({d_{A}d_{B}} ),
\label{Eq.18}
    \end{split}
    \end{align}
herein, $S(\hat{Q} \left | B \right  )$ denotes the conditional von Neumann  entropy of the post-measurement state  $\hat{\rho } _{\hat{Q }B}$ after measuring observable $\hat{Q }$ on subsystem $A$, where $S( \hat{\rho}_{B}  )=-\operatorname{Tr}(\hat{\rho}_{B}\log_{2}{\hat{\rho}}_{B} )$ is the von Neumann entropy of subsystem
$B$. The  parameter
 $c=\mathrm{max}_{i,j} \left |\left \langle \hat{q}_{i}  | \hat{r}_{j} \right  \rangle   \right | ^{2}$ quantifies the maximum overlap between eigenstates $\left | \hat{q}_{i}    \right \rangle$ of $\hat{Q }$ and $\left | \hat{r}_{j}    \right \rangle$ of $\hat{R }$. For a two-qubit system (where  $A$ and  $B$ are in  two-dimensional  Hilbert space),  Eq. (\ref{Eq.18}) is reduced to: $S(\hat{Q} \left | B \right  )+S(\hat{R} \left | B \right  )+S( \hat{\rho}_{B}  )+P_{\text{vn}}( \hat{\rho}_{AB}  )+C_{\text{re}}(\hat{\rho}_{AB}) \geqslant3$.
 \begin{figure}
\begin{minipage}{0.5\textwidth}
\centering
\subfigure{\includegraphics[height=3.78cm]{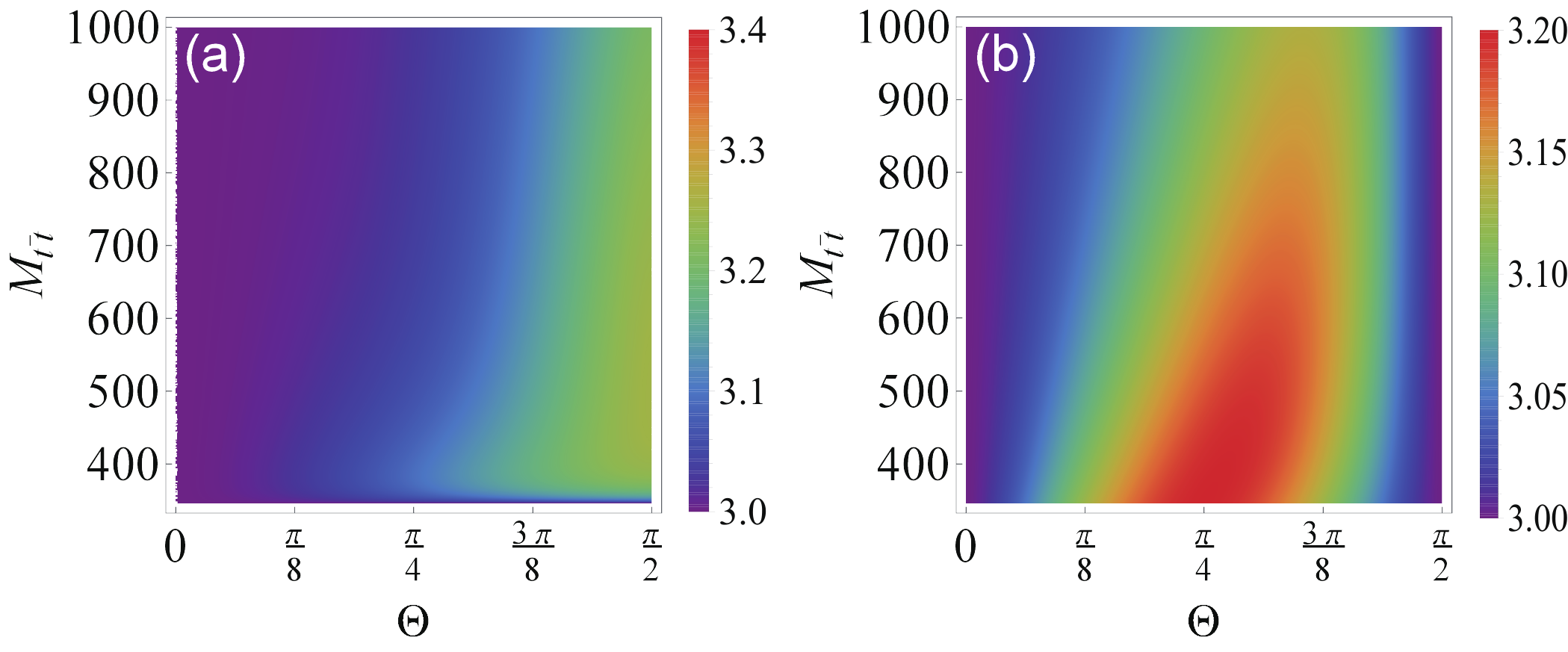}}
\end{minipage}\hfill
    \caption{ The left-hand side of the intrinsic relation as a function of the invariant mass $M_{t\bar{t} }$ and the production angle $\Theta$ in a $t\bar{t}$ pairs. (a): $gg \to t\bar{t}$. (b): $q\bar{q} \to t\bar{t}$.}
    \label{f6}
  \end{figure}

\begin{figure}
\begin{minipage}{0.5\textwidth}
\centering
\subfigure{\includegraphics[height=7.4cm]{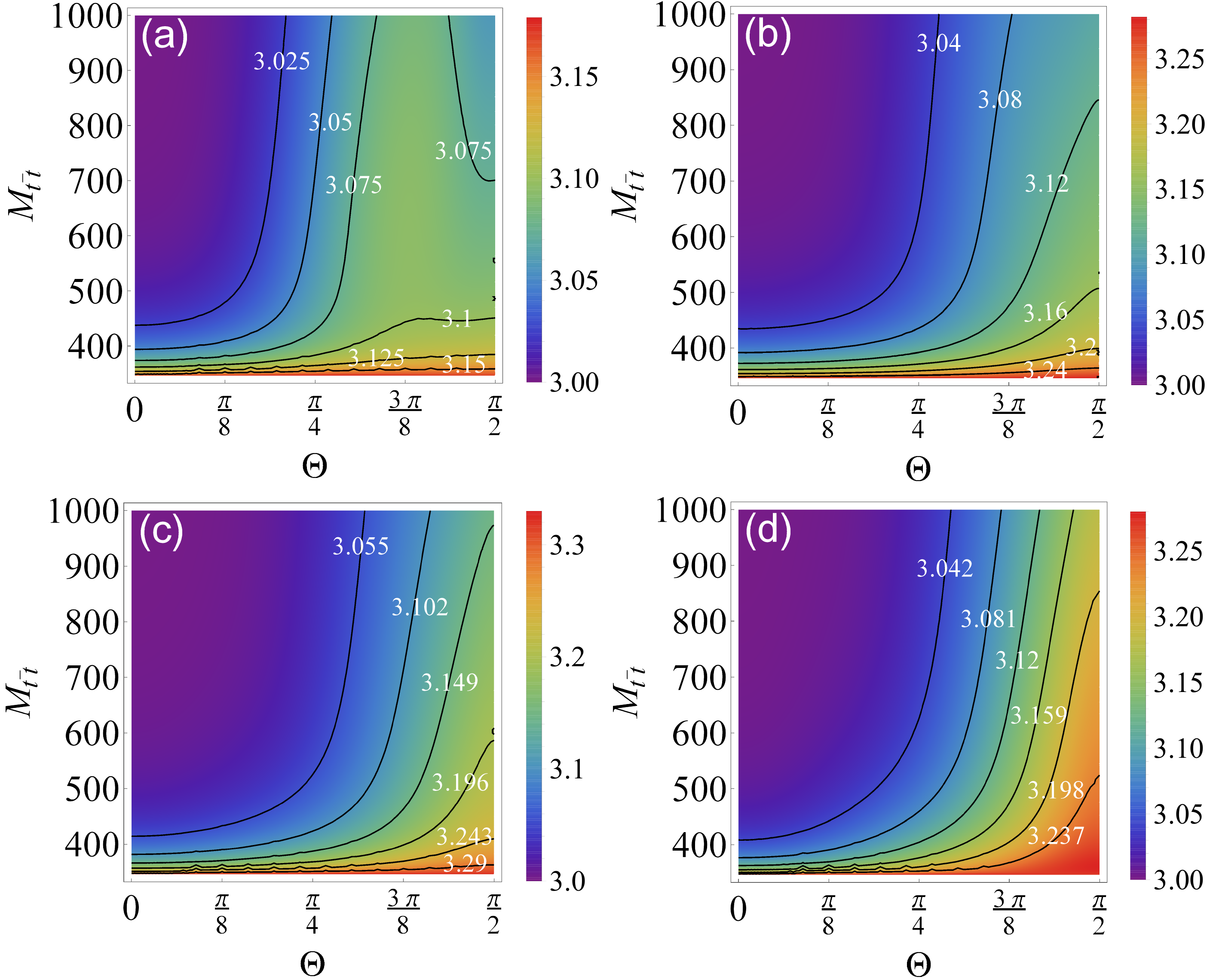}}
\end{minipage}\hfill
    \caption{The left-hand side of the intrinsic relation in $t\bar{t}$ production for mixed gluons $(gg)$ and quarks $(q\bar{q})$ initial state with varying gluons   probability $\mathcal{W} _{gg}$ varies:
    (a): $\mathcal{W} _{gg}=0.2$; (b):  $\mathcal{W} _{gg}=0.4$; (c):  $\mathcal{W} _{gg}=0.6$; (d):  $\mathcal{W} _{gg}=0.8$.}
    \label{f7}
  \end{figure}

\begin{figure}
\begin{minipage}{0.5\textwidth}
\centering
\subfigure{\includegraphics[height=5.6cm]{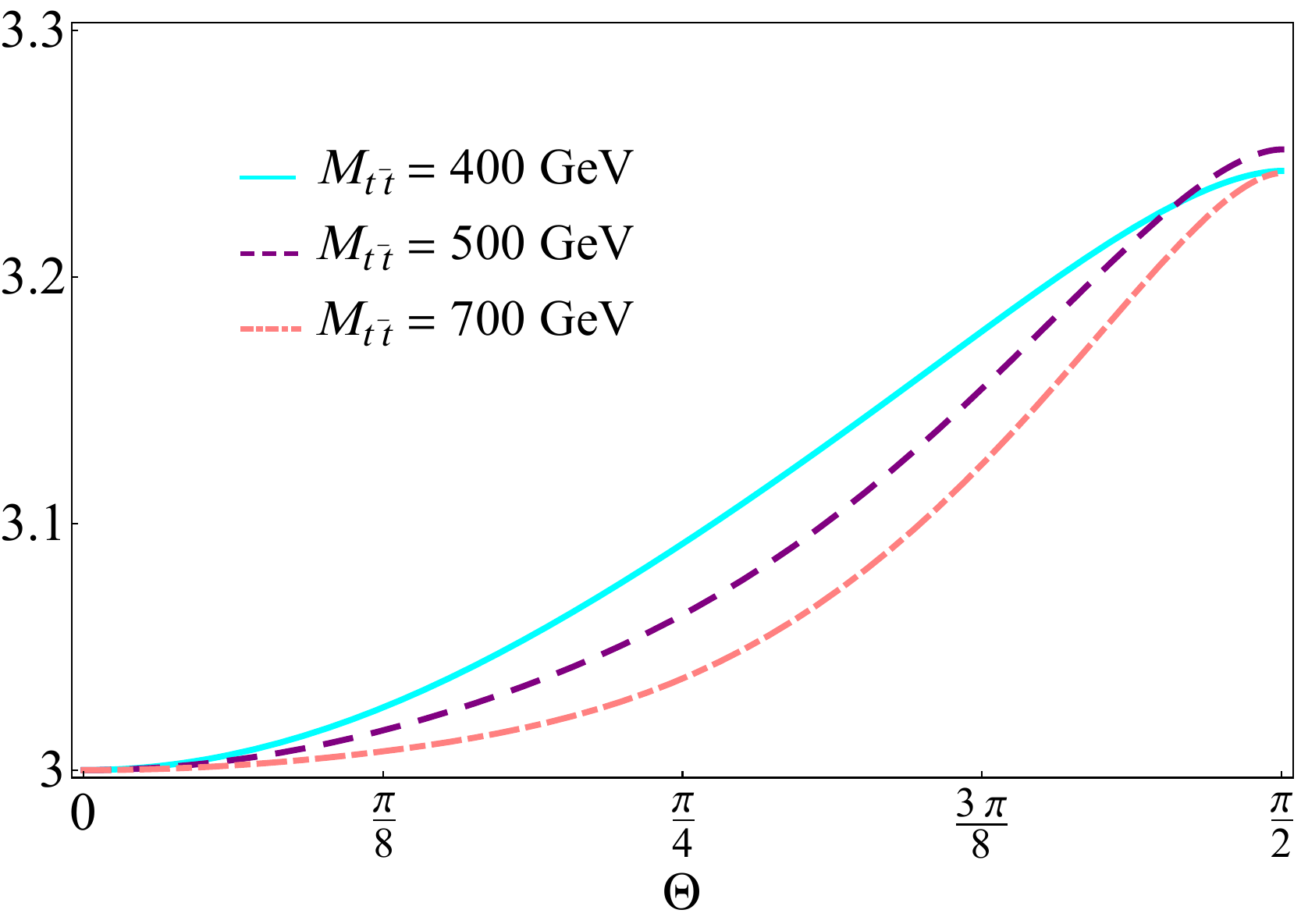}}
\end{minipage}\hfill
    \caption{The left-hand side of the intrinsic relation is plotted as a function of the production angle $\Theta$ for fixed invariant masses $M_{t\bar{t}} = 400$, $500$, and $700$ GeV ($gg \to t\bar{t}$). Cyan line represents the invariant masses $M_{t\bar{t}} = 400$, purple dash-dotted line shows $M_{t\bar{t}} = 500$ and pink dash-dotted line  indicates $M_{t\bar{t}} = 700$.}
    \label{f8}
  \end{figure}

In Fig. \ref{f6}(a), we show the left hand side of the intrinsic relation for $t\bar{t}$ production via gluon fusion ($gg \to t\bar{t}$), illustrating its dependence on the production angle  $\Theta$ and the invariant mass $M_{t\bar{t} }$.  It is observed that for a fixed  $M_{t\bar{t} }$, the left hand side of the intrinsic relation increases monotonically with $\Theta$. In Fig. \ref{f6}(b), the left hand side of the intrinsic relation reaches larger values, especially around  $\Theta\approx\pi /4$.

\begin{figure*}

\centering
\subfigure{\includegraphics[height=3.81cm]{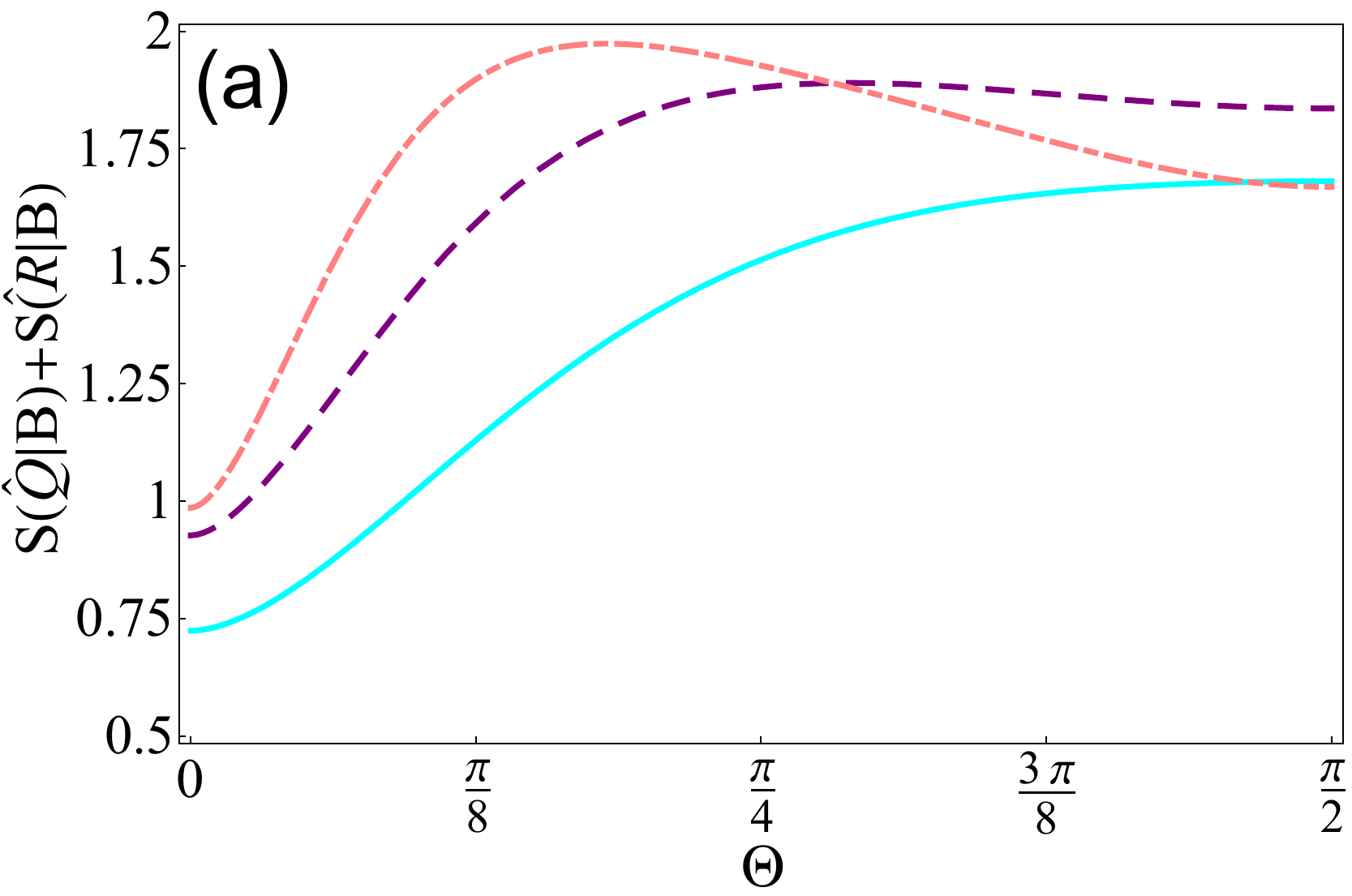}}\
\hspace{0.15cm}         \
\subfigure{\includegraphics[height=3.81cm]{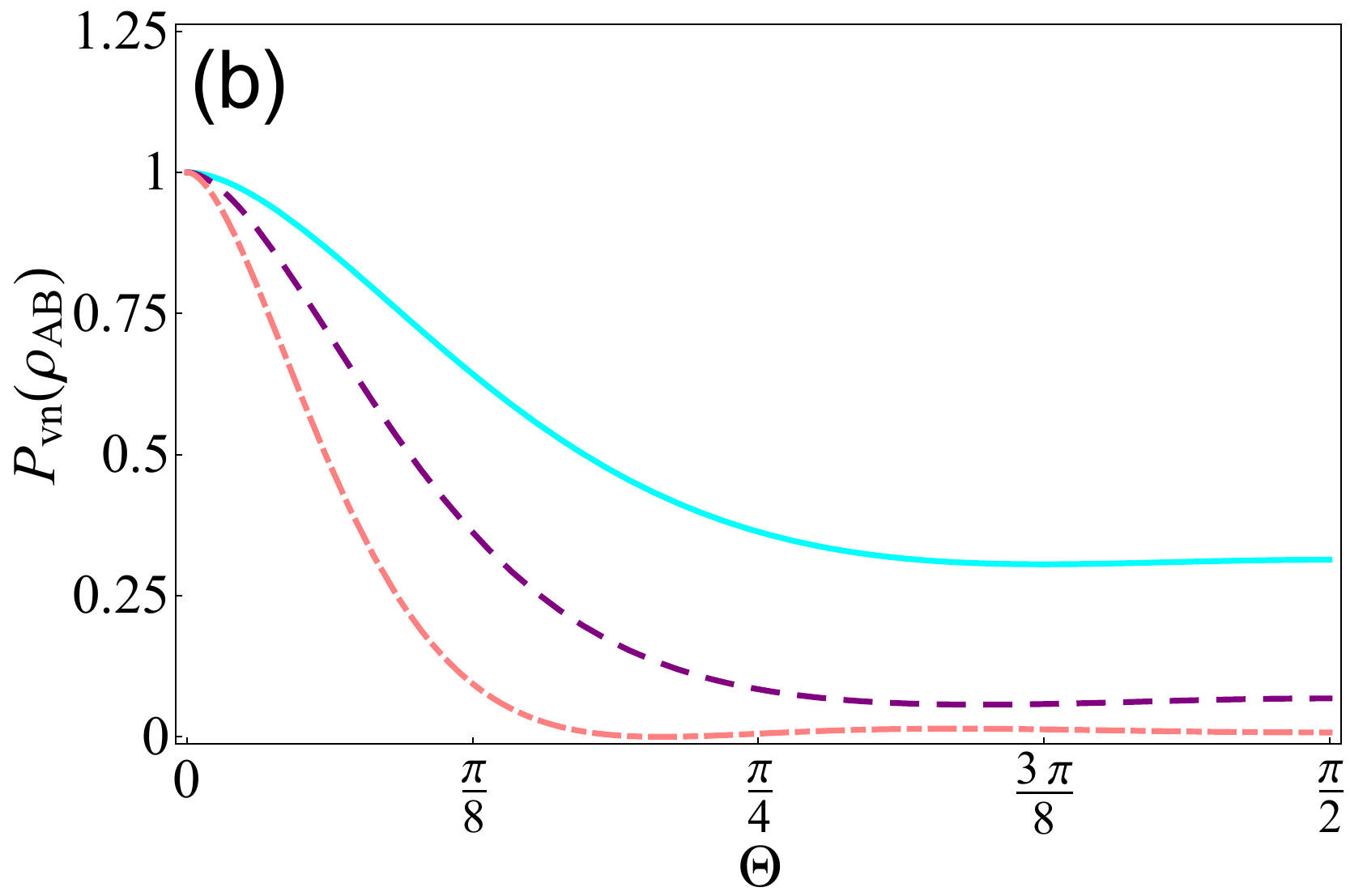}}\       \
\hspace{0.15cm}
\subfigure{\includegraphics[height=3.81cm]{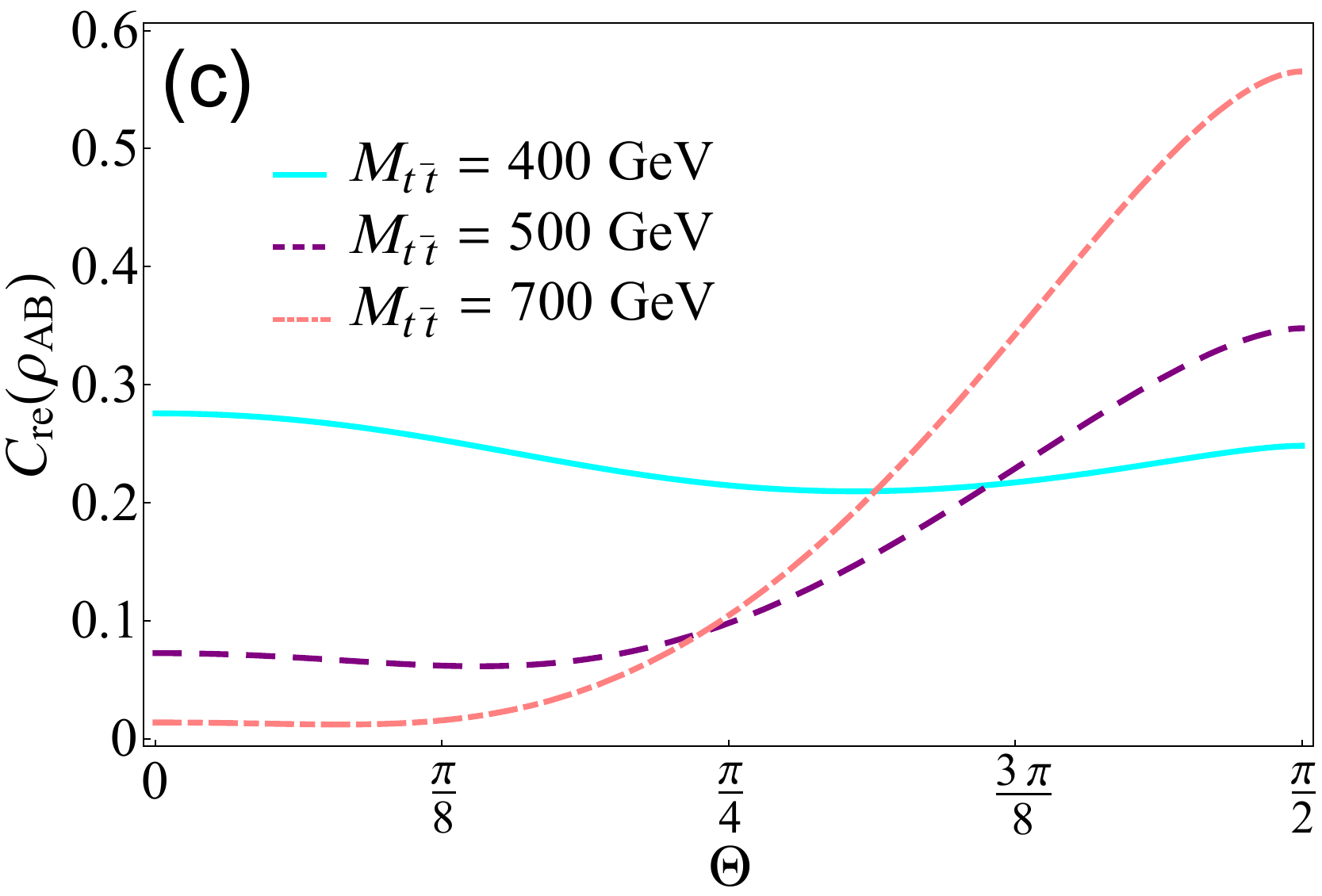}}
\caption{The left-hand side of the intrinsic relation $S(\hat{Q} \left | B \right  )+S(\hat{R} \left | B \right  )+S( \hat{\rho}_{B}  )+P_{\text{vn}}( \hat{\rho}_{AB}  )+C_{\text{re}}(\hat{\rho}_{AB})$ is plotted as a function of the production angle $\Theta$ for fixed invariant masses $M_{t\bar{t}} = 400$, $500$, and $700$ GeV ($gg \to t\bar{t}$). Cyan line represents the invariant masses $M_{t\bar{t}} = 400$, purple dash-dotted line shows $M_{t\bar{t}} = 500$ and pink dash-dotted line  indicates $M_{t\bar{t}} = 700$. (a): The sum of conditional entropies $S(\hat{Q} \left | B \right  )+S(\hat{R} \left | B \right  )$; (b): Predictability measure $P_{\text{vn}}(\hat{\rho} _{AB} )$; (c): REC $C_{\text{re}}(\hat{\rho} _{AB} )$.}
\label{f9}
\end{figure*}

Fig. \ref{f7} plots the left-hand side of the  intrinsic relation in terms of von Neumann  entropy as a function of   production angle $\Theta$  and invariant mass $M_{t\bar{t} }$.  The inequality $S(\hat{Q} \left | B \right  )+S(\hat{R} \left | B \right  )+S( \hat{\rho}_{B}  )+ P_{\text{vn}}( \hat{\rho}_{AB}  )+C_{\text{re}}(\hat{\rho}_{AB}) \geqslant3$ holds for all initial-state mixtures ${W} _{gg}$ of the gluon ($gg$) and  quark ($q\bar{q}$) channels. Panels (a)-(d) reveal that increasing the gluon probability ${W} _{gg}$ enhances the  value of the left-hand side of the  intrinsic relation.   Notably, Fig. \ref{f7}(d)  shows  the maximum value occurs at small invariant mass $M_{t\bar{t} }$ and large  production angle $\Theta$.

To further illustrate the dependence on the invariant mass, we present   the left-hand side of the intrinsic relation as a function of the production angle $\Theta$ ($gg \to t\bar{t}$) for fixed  $M_{t\bar{t}} $ values of $= 400$, $500$, and $700$ GeV  in Fig. \ref{f8}. As shown in the figure, the left-hand side of the intrinsic relation increases monotonically with $\Theta$ for each fixed mass. At fixed production angle $\Theta$, the value exhibits a clear dependence on the invariant mass. As shown in the figure, increasing $M_{t\bar{t} }$ from 500 GeV to 700 GeV does  lead to a decrease for the left-hand side of the intrinsic relation.

Fig. \ref{f9} presents a detailed analysis of the intrinsic relation in $t\bar{t}$ production via the gluon fusion channel ($gg \to t\bar{t}$).  The left-hand side of the intrinsic relation $S(\hat{Q} \left | B \right  )+S(\hat{R} \left | B \right  )+S( \hat{\rho}_{B}  )+P_{\text{vn}}( \hat{\rho}_{AB}  )+C_{\text{re}}(\hat{\rho}_{AB})$  is plotted as a function of the production angle $\Theta$  for three fixed invariant masses: $M_{t\bar{t}} = 400$, $500$, and $700$ GeV. The three subfigures correspond to the individual contributions: (a) the sum of conditional entropies $S(\hat{Q} \left | B \right  )+S(\hat{R} \left | B \right  )$, (b) the predictability measure $P_{\text{vn}}( \hat{\rho}_{AB}  )$, and (c) REC $C_{\text{re}}(\hat{\rho}_{AB})$. Fig. \ref{f9}(b) shows that the predictability $P_{\text{vn}}( \hat{\rho}_{AB}  )$ exhibits a decrease followed by stabilization with increasing $\Theta$ for fixed invariant masses, indicating that the quantum state  becomes less predictable at larger production angles $\Theta$. For invariant masses of $500$ and $700$ GeV, Fig. \ref{f9}(c) shows that the REC $C_{\text{re}}(\hat{\rho}_{AB})$ remains  relatively stable for small  $\Theta$ before rising  as $\Theta$ approaches  $\pi /4$. This demonstrates that quantum coherence is significantly enhanced in the large-angle regime, with the effect being more pronounced at higher energy scales.

\section{Discussion and Summary} \label{sect7}
This work presents the systematic investigation of quantum correlations in top-quark pairs ($t\bar{t}$) production, establishing a novel connection between quantum information theory and high-energy physics. By analyzing $t\bar{t}$
production through QMI, REC, CCR, and their intrinsic relations, we have uncovered several key insights: in the gluon-fusion channel ($gg \to t\bar{t}$), QMI peaks at an invariant mass $M_{t\bar{t} }\approx 346$ GeV, demonstrating strong dependence on both the invariant mass $M_{t\bar{t} }$ and the production angle $\Theta$. For $q\bar{q} \to t\bar{t}$ process, the larger the invariant mass $M_{t\bar{t} }$ and  production angle $\Theta$, the stronger the correlation (QMI) between the two subsystems. When the invariant mass $M_{t\bar{t} }$ is around $346$ GeV, the REC for $gg \to t\bar{t}$ process decreases with the increasing $\Theta$, while the REC for  $q\bar{q}\to t\bar{t}$ process maintains at a relatively large value in most region with $\Theta$ increases.  The analysis shows that the sum of QMI and conditional entropy remains conserved at unity  (i.e., $I_{A:B}(\hat{\rho} _{AB} )+S_{A|B } (\hat{\rho} _{AB} )=1$) in $t\bar{t}$ production, indicating that subsystem $A$ undergoes complete decoherence, collapsing into a classical diagonal state  phenomenon likely driven by environmental interactions that fully erase its quantum coherence. Furthermore, we have rigorously derived the intrinsic relation connecting uncertainty, REC and predictability in terms of  von Neumann entropy. Notably, the left-hand side of this inequality grows with $\mathcal{W }_{gg}$. These findings highlight the profound quantum-informational structure inherent to high-energy particle collisions, offering a unified perspective that bridges quantum field theory and quantum information science. The framework developed here can be extended to other heavy-quark systems, opening new avenues for exploring quantum correlations in beyond-Standard-Model physics and collider experiments.

\begin{acknowledgements}
  This work was supported by the National Science Foundation of China (Grant Nos. 12475009, 12075001, and 62471001), Anhui Provincial Key Research and Development Plan (Grant No. 2022b13020004), Anhui Province Science and Technology Innovation Project (Grant No. 202423r06050004), Anhui Provincial Natural Science Foundation (Grant no. 2508085ZD001), Anhui Provincial Department of Industry and Information Technology (Grant no. JB24044), and Anhui Provincial University Scientific Research Major Project (Grant No. 2024AH040008).
\end{acknowledgements}

\appendix  %
\begin{widetext}
\section{}\label{app:rec}
Based on Eqs. (\ref{Eq.14}) and  (\ref{Eq.15}), the analytical expressions of QMI and REC for the production of  $t\bar{t}$ pairs  are given by
\begin{align}
\begin{split}
&I_{A:B}(\hat{\rho} _{AB} )\\
&=2+\frac{1}{4\tilde{A}^{gg} \tilde{A}^{q\bar{q}  }\log_{}{2}  }([\mathcal{W }_{gg}\tilde{A}^{q\bar{q}}(\tilde{A}^{gg}+\tilde{C}_{rr}^{gg}-\tilde{C}_{nn}^{gg}+\tilde{C}_{kk}^{gg})+\mathcal{W }_{q\bar{q}}\tilde{A}^{gg}(\tilde{A}^{q\bar{q}}+\tilde{C}_{rr}^{q\bar{q}}-\tilde{C}_{nn}^{q\bar{q}}+\tilde{C}_{kk}^{q\bar{q}})]\\
&\times\log_{}{[\frac{\mathcal{W }_{gg}(\tilde{A}^{gg }+\tilde{C}_{rr}^{gg}-\tilde{C}_{nn}^{gg}+\tilde{C}_{kk}^{gg})}{4\tilde{A}^{gg }} -\frac{\mathcal{W }_{q\bar{q}}(\tilde{A}^{q\bar{q} }+\tilde{C}_{rr}^{q\bar{q}}-\tilde{C}_{nn}^{q\bar{q}}+\tilde{C}_{kk}^{q\bar{q}})}{4\tilde{A}^{q\bar{q} }} ]}\\
&+ [\tilde{A}^{gg}\tilde{A}^{q\bar{q}}-\mathcal{W }_{gg}\tilde{A}^{q\bar{q}}(\tilde{C}_{rr}^{gg}+\tilde{C}_{nn}^{gg}+\tilde{C}_{kk}^{gg})-\mathcal{W }_{q\bar{q}}\tilde{A}^{gg}(\tilde{C}_{rr}^{q\bar{q}}+\tilde{C}_{nn}^{q\bar{q}}+\tilde{C}_{kk}^{q\bar{q}})]\\
&\times\log_{}{[\frac{1}{4} -\frac{\mathcal{W }_{gg}(\tilde{C}_{rr}^{gg}+\tilde{C}_{nn}^{gg}+\tilde{C}_{kk}^{gg})}{4\tilde{A}^{gg }} -\frac{\mathcal{W }_{q\bar{q}}(\tilde{C}_{rr}^{q\bar{q}}+\tilde{C}_{nn}^{q\bar{q}}+\tilde{C}_{kk}^{q\bar{q}})}{4\tilde{A}^{q\bar{q} }} ]}\\
&+[\mathcal{W }_{q\bar{q}}\tilde{A}^{gg }(\tilde{A}^{q\bar{q} }+\tilde{C}_{nn}^{q\bar{q}} )+ \mathcal{W }_{gg}\tilde{A}^{q\bar{q} }(\tilde{A}^{gg }+\tilde{C}_{nn}^{gg})-Q]\log_{}[{\frac{\mathcal{W }_{q\bar{q}}\tilde{A}^{gg }(\tilde{A}^{q\bar{q} }+\tilde{C}_{nn}^{q\bar{q}} )+ \mathcal{W }_{gg}\tilde{A}^{q\bar{q} }(\tilde{A}^{gg }+\tilde{C}_{nn}^{gg})-Q}{4\tilde{A}^{gg }\tilde{A}^{q\bar{q} }} }]\\
&+[\mathcal{W }_{q\bar{q}}\tilde{A}^{gg }(\tilde{A}^{q\bar{q} }+\tilde{C}_{nn}^{q\bar{q}} )+ \mathcal{W }_{gg}\tilde{A}^{q\bar{q} }(\tilde{A}^{gg }+\tilde{C}_{nn}^{gg})+Q]\log_{}[{\frac{\mathcal{W }_{q\bar{q}}\tilde{A}^{gg }(\tilde{A}^{q\bar{q} }+\tilde{C}_{nn}^{q\bar{q}} )+ \mathcal{W }_{gg}\tilde{A}^{q\bar{q} }(\tilde{A}^{gg }+\tilde{C}_{nn}^{gg})+Q }{4\tilde{A}^{gg }\tilde{A}^{q\bar{q} }} }])
\label{Eq.19}
    \end{split}
    \end{align}
and

\begin{align}
\begin{split}
&C_{\text{re}}(\hat{\rho}_{AB})\\
&=\frac{1}{4\tilde{A}^{gg} \tilde{A}^{q\bar{q}  }\log_{}{2}  }(4\tilde{A}^{gg} \tilde{A}^{q\bar{q}  }\log_{}{4} +2(\mathcal{W }_{gg}\tilde{A}^{q\bar{q}}\tilde{C}_{kk}^{gg}+\mathcal{W }_{q\bar{q}}\tilde{A}^{gg}\tilde{C}_{kk}^{q\bar{q}}-\tilde{A}^{gg} \tilde{A}^{q\bar{q}  } )\log_{}{[1-\frac{\mathcal{W }_{gg}\tilde{C}_{kk}^{gg}}{\tilde{A}^{gg}}-\frac{\mathcal{W }_{q\bar{q}}\tilde{C}_{kk}^{q\bar{q}}}{\tilde{A}^{q\bar{q}}}  ]}\\
&-2[\mathcal{W }_{gg}\tilde{A}^{q\bar{q}}(\tilde{A}^{gg}+\tilde{C}_{kk}^{gg})+\mathcal{W }_{q\bar{q}}\tilde{A}^{gg}(\tilde{A}^{q\bar{q}}+\tilde{C}_{kk}^{q\bar{q}})]\log_{}{[1+\frac{\mathcal{W }_{gg}\tilde{C}_{kk}^{gg}}{\tilde{A}^{gg}}+\frac{\mathcal{W }_{q\bar{q}}\tilde{C}_{kk}^{q\bar{q}}}{\tilde{A}^{q\bar{q}}}]}\\
&+[\mathcal{W }_{gg}\tilde{A}^{q\bar{q}}(\tilde{A}^{gg}+\tilde{C}_{rr}^{gg}-\tilde{C}_{nn}^{gg}+\tilde{C}_{kk}^{gg})+\mathcal{W }_{q\bar{q}}\tilde{A}^{gg}(\tilde{A}^{q\bar{q}}+\tilde{C}_{rr}^{q\bar{q}}-\tilde{C}_{nn}^{q\bar{q}}+\tilde{C}_{kk}^{q\bar{q}})]\\
&\times\log_{}{[\frac{\mathcal{W }_{gg}(\tilde{A}^{gg }+\tilde{C}_{rr}^{gg}-\tilde{C}_{nn}^{gg}+\tilde{C}_{kk}^{gg})}{4\tilde{A}^{gg }} -\frac{\mathcal{W }_{q\bar{q}}(\tilde{A}^{q\bar{q} }+\tilde{C}_{rr}^{q\bar{q}}-\tilde{C}_{nn}^{q\bar{q}}+\tilde{C}_{kk}^{q\bar{q}})}{4\tilde{A}^{q\bar{q} }} ]}\\
&+ [\tilde{A}^{gg}\tilde{A}^{q\bar{q}}-\mathcal{W }_{gg}\tilde{A}^{q\bar{q}}(\tilde{C}_{rr}^{gg}+\tilde{C}_{nn}^{gg}+\tilde{C}_{kk}^{gg})-\mathcal{W }_{q\bar{q}}\tilde{A}^{gg}(\tilde{C}_{rr}^{q\bar{q}}+\tilde{C}_{nn}^{q\bar{q}}+\tilde{C}_{kk}^{q\bar{q}})]\\
&\times\log_{}{[\frac{1}{4} -\frac{\mathcal{W }_{gg}(\tilde{C}_{rr}^{gg}+\tilde{C}_{nn}^{gg}+\tilde{C}_{kk}^{gg})}{4\tilde{A}^{gg }} -\frac{\mathcal{W }_{q\bar{q}}(\tilde{C}_{rr}^{q\bar{q}}+\tilde{C}_{nn}^{q\bar{q}}+\tilde{C}_{kk}^{q\bar{q}})}{4\tilde{A}^{q\bar{q} }} ]}\\
&+[\mathcal{W }_{q\bar{q}}\tilde{A}^{gg }(\tilde{A}^{q\bar{q} }+\tilde{C}_{nn}^{q\bar{q}} )+ \mathcal{W }_{gg}\tilde{A}^{q\bar{q} }(\tilde{A}^{gg }+\tilde{C}_{nn}^{gg})-Q]\log_{}[{\frac{\mathcal{W }_{q\bar{q}}\tilde{A}^{gg }(\tilde{A}^{q\bar{q} }+\tilde{C}_{nn}^{q\bar{q}} )+ \mathcal{W }_{gg}\tilde{A}^{q\bar{q} }(\tilde{A}^{gg }+\tilde{C}_{nn}^{gg})-Q}{4\tilde{A}^{gg }\tilde{A}^{q\bar{q} }} }]\\
&+[\mathcal{W }_{q\bar{q}}\tilde{A}^{gg }(\tilde{A}^{q\bar{q} }+\tilde{C}_{nn}^{q\bar{q}} )+ \mathcal{W }_{gg}\tilde{A}^{q\bar{q} }(\tilde{A}^{gg }+\tilde{C}_{nn}^{gg})+Q]\log_{}[{\frac{\mathcal{W }_{q\bar{q}}\tilde{A}^{gg }(\tilde{A}^{q\bar{q} }+\tilde{C}_{nn}^{q\bar{q}} )+ \mathcal{W }_{gg}\tilde{A}^{q\bar{q} }(\tilde{A}^{gg }+\tilde{C}_{nn}^{gg})+Q }{4\tilde{A}^{gg }\tilde{A}^{q\bar{q} }} }]),
\label{Eq.20}
    \end{split}
    \end{align}
respectively, where $f_{1}^{gg}=(\mathcal{W }_{gg})^{2}(\tilde{A}^{q\bar{q} })^{2} [4(\tilde{C}_{rk}^{gg})^{2}+(\tilde{C}_{rr}^{gg}-\tilde{C}_{kk}^{gg})^{2}]$, $f_{1}^{q\bar{q}}=(\mathcal{W }_{q\bar{q}})^{2}(\tilde{A}^{gg})^{2}[4(\tilde{C}_{rk}^{q\bar{q}})^{2}+(\tilde{C}_{rr}^{q\bar{q}}-\tilde{C}_{kk}^{q\bar{q}})^{2}]$ and $Q=\sqrt{f_{1}^{gg}+f_{1}^{q\bar{q}}
+2\mathcal{W }_{gg}\mathcal{W }_{q\bar{q}}\tilde{A}^{gg }\tilde{A}^{q\bar{q} }[4\tilde{C}_{rk}^{gg}\tilde{C}_{rk}^{q\bar{q}}+(\tilde{C}_{rr}^{q\bar{q}}-\tilde{C}_{kk}^{q\bar{q}})(\tilde{C}_{rr}^{gg}-\tilde{C}_{kk}^{gg})]
}$.
For the production of  $t\bar{t}$ pairs via gluons (quarks), $\mathcal{W }_{gg}=1 (\mathcal{W }_{q\bar{q}}=1)$.
\end{widetext}

%

\end{document}